%% file: Template.tex
\documentclass{article}
\usepackage{amsmath,graphicx}
\usepackage[preprint]{spconf}
\usepackage{amssymb}
\usepackage{bm}

\usepackage{siunitx}
\usepackage{tikz}
\usepackage{tikz, tikzscale}
\usetikzlibrary{matrix,chains,positioning,decorations.pathreplacing,arrows,fadings}

\usepackage{pgfplotstable}
\usepackage{pgfplots}
\usepgfplotslibrary{units}
\usetikzlibrary{pgfplots.statistics, pgfplots.colorbrewer}
\usepackage{siunitx}
\pgfplotsset{compat=1.9}
\makeatletter
\pgfplotsset{
    unit code/.code 2 args=
    \begingroup
    \protected@edef\x{\endgroup\si{#2}}\x
}

\usepackage{cite}
\def\x{{\mathbf x}}

\title{Reconstruction of sound field through diffusion models}
\name{\begin{minipage}{\textwidth}\centering Federico Miotello, Luca Comanducci, Mirco Pezzoli,\\Alberto Bernardini, Fabio Antonacci, Augusto Sarti\end{minipage}\thanks{This work has been funded by "REPERTORIUM project. Grant agreement number 101095065. Horizon Europe. Cluster II. Culture, Creativity and Inclusive Society. Call HORIZON-CL2-2022-HERITAGE-01-02."}\thanks{This work was supported by the Italian Ministry of University and Research (MUR) under the National Recovery and Resilience Plan (NRRP), and by the European Union (EU) under the NextGenerationEU project.}\address{Dipartimento di Elettronica, Informazione e Bioingegneria - Politecnico di Milano, Milan, Italy}}
\toappear{\begin{minipage}{\textwidth}\footnotesize© 20XX IEEE. Personal use of this material is permitted. Permission from IEEE must be obtained for all other uses, in any current or future media, including reprinting/republishing this material for advertising or promotional purposes, creating new collective works, for resale or redistribution to servers or lists, or reuse of any copyrighted component of this work in other works.\end{minipage}}
\begin{document}
%
\maketitle
\begin{abstract}
Reconstructing the sound field in a room is an important task for several applications, such as sound control and augmented (AR) or virtual reality (VR). 
In this paper, we propose a data-driven generative model for reconstructing the magnitude of acoustic fields in rooms with a focus on the modal frequency range.
We introduce, for the first time, the use of a conditional Denoising Diffusion Probabilistic Model (DDPM) trained in order to reconstruct the sound field (SF-Diff) over an extended domain. 
The architecture is devised in order to be conditioned on a set of limited available measurements at different frequencies and generate the sound field in target, unknown, locations.
The results show that SF-Diff is able to provide accurate reconstructions.
We conduct a comparative analysis with two state-of-the-art baseline methods, one relying on kernel interpolation and the other on deep learning.

\end{abstract}
%
\input{introduction}
\input{problem}
\input{diffusion_rtf}
\input{results}

\input{conclusion}
\bibliographystyle{ieeetr}
{\footnotesize
\bibliography{strings,refs}}

\end{document}

%% file: introduction.tex
\section{Introduction}\label{sec:introduction}
Sound field reconstruction is a relevant problem in the field of acoustic signal processing, especially when considering the modal frequency range, due to its importance in applications such as sound field control and room compensation~\cite{schmid2021spatial}.
Applications include, for example, the navigation of acoustic scenes \cite{tylka2020fundamentals}.
The goal is to estimate the acoustic field over an extended area starting from the information provided by a limited set of sensors.
In the literature, several techniques tackling the reconstruction of sound field can be found.
Most solutions rely on either a parametric description \cite{thiergart2013geometry,pezzoli2020parametric,mccormack2022parametric,pezzoli2018reconstruction} of the acoustic scene, or on models based on the solutions of the wave equations \cite{ueno2017sound, ueno2018kernel} including plane waves \cite{jin2015theory}, spherical waves \cite{pezzoli2022sparsity, borra2019soundfield,fahim2017sound} or equivalent sources \cite{koyama2019sparse,daminano2021sound}. 
In \cite{ueno2017sound, ueno2018kernel} and variations \cite{ribeiro2022region,ribeiro2023kernel}, the reconstruction of the acoustic field is achieved using a kernel-interpolation based approach which exploit the solution of the Helmholtz equation as a physically motivated kernel.

A different class of techniques is based on deep learning, which has been widely applied in the field of acoustics \cite{bainco2019acousticdeepreview, olivieri2021physics}. 
The main advantage of deep learning solutions is to adopt more sparse and irregular microphone array setups for the sound field reconstruction.
The first learning-based approach was proposed in~\cite{lluis2020sound} and consisted of a U-Net architecture, which was applied in order to reconstruct the magnitude of the sound field with an approach similar to image inpainting.
Similarly, in \cite{pezzoli2022deep}, the authors proposed a deep-prior approach to RIR reconstruction following the deep prior paradigm introduced for image inpainting \cite{ulyanov2018deep}.
This approach assumes that the structure of the CNN introduces an implicit prior regularizing the estimation of RIRs.
Other solutions instead \cite{pezzoli2023implicit, karakonstantis2023generative}, rely on the physical equation governing the sound propagation i.e., the wave equation as an alternative approach for improving the reconstruction.
In \cite{pezzoli2023implicit}, the authors introduce a physics-informed neural newtwork for the reconstruction of acoustic fields in the time domain.
The model employed a SIREN architecture \cite{sitzmann2020implicit} trained with a physics-informed loss function including the wave equation. 
The model has been tested on time-domain sound fields captured through a uniform linear microphone array.
Although effective, this techniques require to perform per-element training. 

More recently, also generative models such as Generative Adversarial Networks (GANs) have been applied to sound field reconstruction problems~\cite{fernandez2023generative}. 
In particular, in \cite{fernandez2023generative} three different generative models based on GANs are considered: a compressive sensing model, a conditional-GAN and a High-fidelity-GAN.
The results in \cite{fernandez2023generative} proved that generative neural networks provide an effective approach for the reconstruction of acoustic field.
Nonetheless, among generative models, Denoising Diffusion Probabilistic Models (DDPMs)~\cite{ho2020denoising} have recently gained interest, due to their enhanced synthesis capabilities and more stable training process with respect to GANs in different tasks \cite{rombach2022high, moliner2023solving}. 
As such, they have been applied to a variety of sound-related problems such as speech enhancement~\cite{yen2023cold}, speech super-resolution~\cite{yu2023conditioning}, vocoders~\cite{takahashi2023hierarchical} and audio inverse problems in general~\cite{moliner2023solving}.

In this paper, motivated by the superior performance of DDPMs in different fields, we propose a DDPMs approach for the reconstruction of room transfer functions (RTFs).
We consider the sound field reconstruction as an image-to-image translation problem, where DDPMs have already been succesfully applied~\cite{saharia2022palette}.
Specifically, we focus on the sound field measured on a 2D plane using irregular microphone arrays. 
We give as input to the DDPM the magnitude of the computed sound field and inject noise where no microphones are deployed.
Moreover, we condition the model on the embedding that encodes a considered frequency.
Through a simulation campaign, we compare the performance of the proposed method with a kernel interpolation-based signal processing technique~\cite{ueno2018kernel} and a learning-based approach \cite{lluis2020sound}, for configurations with different numbers of microphones, and demonstrate the benefits of applying DDPMs to the sound field reconstruction task.

The rest of the paper is organized as follows: in Sec.~\ref{sec:problem_formulation} we present the adopted data model and formalize the sound field reconstruction problem. In Sec.~\ref{sec:solution} we present the conditional DDPM for sound field reconstruction, while in Sec.~\ref{sec:results} we present results aimed at demonstrating the effectiveness of the proposed technique. Finally, in Sec.~\ref{sec:conclusion}, we draw some conclusions.

%% file: problem.tex
\section{Problem Formulation}\label{sec:problem_formulation}
\subsection{Data model}
Following the approach proposed in \cite{lluis2020sound}, let us consider a three-dimensional rectangular room $\mathcal{R} = [0, l_x] \times [0, l_y] \times [0, l_z]$, where $\{l_x, l_y, l_x\} \in \mathbb{R}^3_{>0}$ denote the length, width, and height of the room, respectively.
The complex-valued frequency-domain sound field in position $\mathbf{r} \in \mathcal{R}$, can be computed using the Fourier transform as
\begin{equation}\label{eq:fourier_transform}
    P(\mathbf{r}, \omega) = \int_\mathbb{R} p(\mathbf{r}, t) e^{-j\omega t} \,dt,
\end{equation}
where $\omega \in \mathbb{R}$ is the angular frequency, and $p(\mathbf{r}, t)$ denotes the acoustic sound field measured at position $\mathbf{r}$ at time $t$.
Moreover, for the purpose of this article, we focus on the magnitude $|P(\mathbf{r}, \omega)|$ of the sound fields.
In practice, room $\mathcal{R}$ is sampled using a regular rectangular grid $\mathcal{D}_o$, with fixed height $z_o \in [0, l_z]$ and defined as
\begin{equation}
    \mathcal{D}_o = \left\{ \left( i \frac{l_x}{I-1}, j\frac{l_y}{J-1}, z_o \right)  \right\}_{i,j},
\end{equation}
where $i = 0, 1, \dots, I-1$, and $j = 0, 1, \dots, J-1$ are the indexes of the sampled points in the grid, with $I,J \geq 2$.

\subsection{Problem definition}
We assume that a limited subset of room measurements $\mathcal{S}_o \subseteq \mathcal{D}_o$ are available.
Sound field magnitude reconstruction can be defined as the problem of recovering the missing data $\{|P(\mathbf{r}, \omega)|\}_{\mathbf{r} \in \mathcal{D}_o \setminus \mathcal{S}_o}$, by exploiting the information in the available observations $\{|P(\mathbf{r}, \omega)|\}_{\mathbf{r} \in \mathcal{S}_o}$ obtained using a limited number of irregularly deployed microphones.

Various techniques have been proposed in the literature to address sound field reconstruction from an under-sampled measurement set \cite{ueno2017sound, pezzoli2023implicit, lluis2020sound, daminano2021sound, karakonstantis2023generative}.
In general, this task can be interpreted in the framework of inverse
problems, and a solution to the following minimization problem
\begin{equation}\label{eq:inverse_problem}
\begin{aligned}
 \bm{\theta}^* &= \underset{{\bm{\theta}}}{\text{argmin}}\,\,J\left(\bm{\theta}\right) = \\
 &E \left( f_{\bm{\theta}}(\{|P(\mathbf{r}, \omega)|\}_{\mathbf{r} \in \mathcal{S}_o}), \{|P(\mathbf{r}, \omega)|\}_{\mathbf{r} \in \mathcal{D}_o \setminus \mathcal{S}_o} \right),
\end{aligned}
\end{equation}
where $f_{\bm{\theta}}(\{|P(\mathbf{r}, \omega)|\}_{\mathbf{r} \in \mathcal{S}_o})$ is a function that generates the estimated sound field using parameters $\bm{\theta}$ having access to available measurements, and $E(\cdot)$ is a data-fidelity term, e.g., the mean squared error (MSE), between the estimated data and the observations.
It is worth noting that in \eqref{eq:inverse_problem}, the evaluation of the reconstruction error is performed in the observed locations $\{\mathbf{r}\} \in \mathcal{S}_o$. 
However, $f$ must be able to provide a meaningful estimate also in locations different from the available ones, i.e., $\{\mathbf{r}\} \in \mathcal{D}_o \setminus \mathcal{S}_o$. 
Therefore, the solution to the ill-posed problem \eqref{eq:inverse_problem} is constrained using regularization strategies. 
Typical techniques include compressed sensing frameworks based on assumptions about the signal model \cite{daminano2021sound}, such as plane and spherical wave expansions \cite{koyama2019sparse} or the RIRs structure \cite{zea2019compressed}, as well as deep learning approaches \cite{lluis2020sound, karakonstantis2023generative}.

%% file: diffusion_rtf.tex
\begin{figure*}[t]
\centering
\begin{minipage}[b]{0.32\textwidth}
  \centering
  \centerline{\input{plot_nmse}}
  \centerline{(a)}
  \vspace{.5em}
\end{minipage}
\hfill
\begin{minipage}[b]{0.32\textwidth}
  \centering
  \centerline{\input{plot_nmse_koyama}}
  \centerline{(b)}
  \vspace{.5em}
\end{minipage}
\hfill
\begin{minipage}[b]{0.32\textwidth}
  \centering
  \centerline{\input{plot_nmse_lluis}}
  \centerline{(c)}
  \vspace{.5em}
\end{minipage}
\ref{mylegend}
\caption{Normalized Mean Squared Error (NMSE) for different number of microphones $m$ measured over the reconstructed magnitude using the proposed method (a), Ueno et al.~\cite{ueno2018kernel} (b) and Lluis et al.~\cite{lluis2020sound} (c).}
\label{fig:nmse}
\end{figure*}
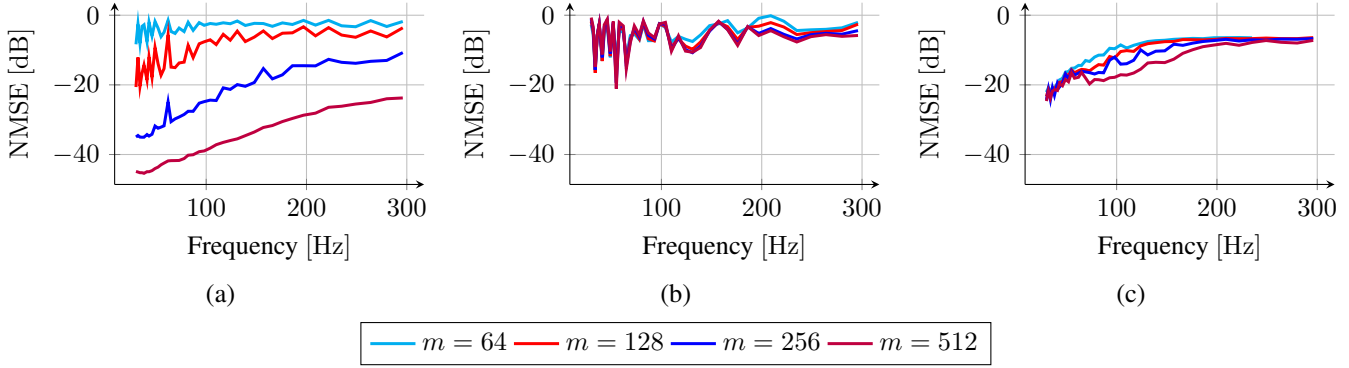%
\section{Proposed Solution}\label{sec:solution}
\subsection{Diffusion Model for RTF reconstruction}
In this work, we aim at solving the sound field reconstruction problem \eqref{eq:inverse_problem} in order to provide an estimate of the magnitude of a sound field as
\begin{equation}
    |\hat{P}(\mathbf{r}, \omega)| = f_{\bm{\theta}}\left(\{|P(\mathbf{r}, \omega)|\}_{\mathbf{r} \in \mathcal{S}_o}\right),
\end{equation}
where the function $f(\cdot)$ represents a neural network.
In particular, we exploit the power of diffusion models, which have recently emerged as the cutting-edge technology in the field of deep learning-based generation, becoming the new state-of-the-art \cite{ho2020denoising}.
In particular, these are quickly replacing Generative Adversarial Networks (GANs) and Variational Autoencoders across various tasks and domains, thanks to their straightforward training process and the higher generation accuracy.
Diffusion models employ an iterative denoising process to transform samples generated from a standard Gaussian distribution into samples that align with an empirical data distribution.
In particular, we employ Palette \cite{saharia2022palette}, a conditional denoising diffusion model of the form $p(\boldsymbol{y} | \boldsymbol{x})$, in which the denoising process is conditioned by an input signal.
For example, in the case of the proposed method, $\boldsymbol{x}$ would be the under-sampled sound field, while $\boldsymbol{y}$ the reconstructed one.
Palette proved to be effective in many image-to-image translation tasks (e.g., inpainting, colorization, uncropping, etc.), outperforming already existing state-of-the-art methods.

The training process of Palette is carried out by optimizing an objective function in the form 
{\small\begin{equation}
    \mathbb{E}_{(\boldsymbol{x}, \boldsymbol{y})} \mathbb{E}_{\boldsymbol{\epsilon} \sim \mathcal{N}(0, I)} \mathbb{E}_\gamma\|f_\theta(\boldsymbol{x}, \underbrace{\sqrt{\gamma} \boldsymbol{y}+\sqrt{1-\gamma} \boldsymbol{\epsilon}}_{\widetilde{\boldsymbol{y}}}, \gamma)-\boldsymbol{\epsilon}\|_2^2,
\end{equation}}
where $\boldsymbol{y}$ is a training output sample, $\boldsymbol{\tilde{y}}$ is its noisy version and $\boldsymbol{\epsilon}$ a noise vector. We train a neural network $f$ with parameters $\bm{\theta}$ to denoise $\boldsymbol{\tilde{y}}$ given $\boldsymbol{y}$ and a noise level indicator $\gamma$ \cite{saharia2022palette}.

\subsection{Architecture}
Palette exploits the U-Net architecture\cite{ronneberger2015u}, which was originally proposed for multimodal medical image segmentation.
In particular, the network has been enhanced with modifications proposed in recent works and is based on the $256 \times 256$ class-conditional U-Net model used in \cite{nichol2021improved}.
However, for the purpose of the task we are addressing and following the approach proposed \cite{saharia2022palette}, the architecture we are employing does not present class-conditioning but rather an additional conditioning of the input data via concatenation \cite{saharia2022image}.

Even though, in principle, the method is able to reconstruct any arbitrary sound field, we focus on the reconstruction of the magnitude of room transfer functions (RTFs), which correspond to the Fourier transform of the impulse response of a room, measured in positions $\mathbf{r} \in \mathcal{D}_o$ and computed as in \eqref{eq:fourier_transform}.
The input to the network is composed by the concatenation of the RTFs magnitude matrix $\mathbf{P}$ and a frequency embedding $\mathbf{F}$.

Similarly to the approach proposed in \cite{saharia2022palette}, in correspondence of the unknown data points in $\mathbf{P}$, the RTF magnitude value is replaced with noise coming from a Gaussian distribution $\mathcal{N}(0,1)$.
The role of $\mathbf{F}$ is to provide the extra conditioning, needed for the diffusion model to consistently learn how to reconstruct the magnitude of a room transfer function at a certain frequency, starting from a noisy version of it.
During the training phase, we restrict the computation of the loss function, i.e., the MSE, only to the the available measures of the RTFs.

%% file: plot_nmse.tex
\begin{tikzpicture}%
\begin{axis}[%
    xlabel={Frequency},%
    x unit = \hertz,%
    ylabel={$\mathrm{NMSE}$},%
    y unit = \decibel, %
    ytick = {-40, -20, 0},
    ymax=0, ymin=-45,
    axis x line=bottom,%
    axis y line=left, %
    grid, %
    height=4cm, %
    width=\columnwidth, %
    enlarge x limits=0.08,%
    enlarge y limits=0.08,%
    style={font=\normalsize},%
    log ticks with fixed point
]

\addplot[mark=, line width=0.4mm, color=cyan] table {results/64.txt};%

\addplot[mark=, line width=0.4mm, color=red] table {results/128.txt};%

\addplot[mark=, line width=0.4mm, color=blue] table {results/256.txt};%

\addplot[mark=, line width=0.4mm, color=purple] table {results/512.txt};%

\end{axis}%
\end{tikzpicture}

%% file: plot_nmse_koyama.tex
\begin{tikzpicture}%
\begin{axis}[%
    xlabel={Frequency},%
    x unit = \hertz,%
    ylabel={$\mathrm{NMSE}$},%
    y unit = \decibel, %
    ytick = {-40, -20, 0},
    ymax=0, ymin=-45,
    axis x line=bottom,%
    axis y line=left, %
    grid=major, %
    height=4cm, %
    width=\columnwidth, %
    enlarge x limits=0.08,%
    enlarge y limits=0.08,%
    style={font=\normalsize},%
    legend to name={mylegend},
    legend columns=4,%
    legend style={at={(0.5,1.1)},anchor=south,font=\normalsize},%
    log ticks with fixed point
]

\addplot[mark=, line width=0.4mm, color=cyan] table {results/nmse_64_koyama.txt};%
\addlegendentry{$m=64$} %

\addplot[mark=, line width=0.4mm, color=red] table {results/nmse_128_koyama.txt};%
\addlegendentry{$m=128$} %

\addplot[mark=, line width=0.4mm, color=blue] table {results/nmse_256_koyama.txt};%
\addlegendentry{$m=256$} %

\addplot[mark=, line width=0.4mm, color=purple] table {results/nmse_512_koyama.txt};%
\addlegendentry{$m=512$} %

\end{axis}%
\end{tikzpicture}

%% file: plot_nmse_lluis.tex
\begin{tikzpicture}%
\begin{axis}[%
    xlabel={Frequency},%
    x unit = \hertz,%
    ylabel={$\mathrm{NMSE}$},%
    y unit = \decibel, %
    ytick = {-40, -20, 0},
    ymax=0, ymin=-45,
    axis x line=bottom,%
    axis y line=left, %
    grid=major, %
    height=4cm, %
    width=\columnwidth, %
    enlarge x limits=0.08,%
    enlarge y limits=0.08,%
    style={font=\normalsize},%
    log ticks with fixed point
]

\addplot[mark=, line width=0.4mm, color=cyan] table {results/nmse_64_lluis.txt};%

\addplot[mark=, line width=0.4mm, color=red] table {results/nmse_128_lluis.txt};%

\addplot[mark=, line width=0.4mm, color=blue] table {results/nmse_256_lluis.txt};%

\addplot[mark=, line width=0.4mm, color=purple] table {results/nmse_512_lluis.txt};%

\end{axis}%
\end{tikzpicture}

%% file: results.tex
\section{Experimental Validation}\label{sec:results}
\begin{figure*}[!htb]
\begin{minipage}[b]{.32\linewidth}
  \centering%
\centerline{\includegraphics[width=.8\columnwidth]{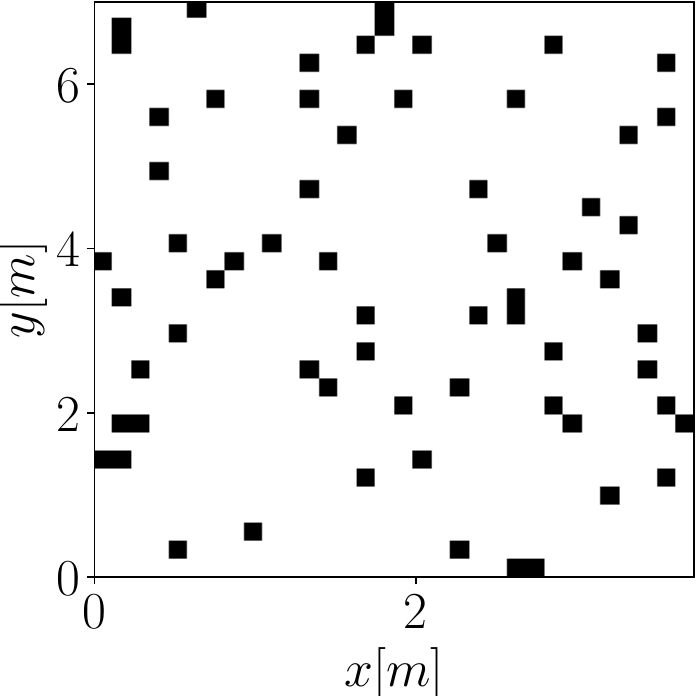}}
  \centerline{(a)}\medskip
\end{minipage}
\hfill
\begin{minipage}[b]{.32\linewidth}
  \centering%
\centerline{\includegraphics[width=.8\columnwidth]{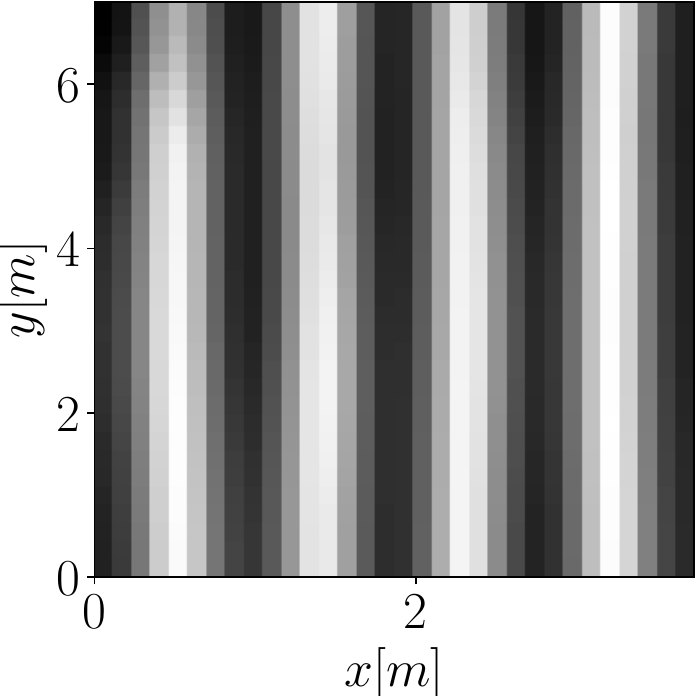}}
  \centerline{(b)}\medskip
\end{minipage}
\hfill
\begin{minipage}[b]{.32\linewidth}
  \centering%
\centerline{\includegraphics[width=.8\columnwidth]{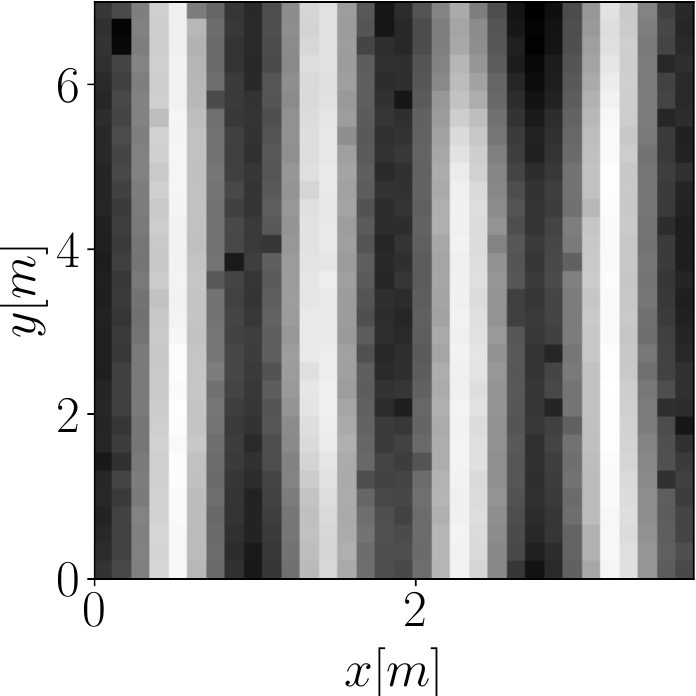}}
  \centerline{(c)}\medskip
\end{minipage}
%
\caption{Magnitude of the sound field in a randomly generated $[3.7~\mathrm{m} \times 7~\mathrm{m} \times26.1~\mathrm{m}]$ room with a $98~\mathrm{Hz}$ active source $\mathbf{s}$ positioned at $[0.9~\mathrm{m}, 0.3~\mathrm{m}, 2.4~\mathrm{m}]^T$, obtained using the proposed method (c), using the $64$ active microphone configuration depicted in (a). Ground truth magnitude is shown in (b).
}
\label{fig:abs_plots}
\end{figure*}

\subsection{Setup}\label{sec:setup}
Following the approach proposed in \cite{lluis2020sound}, we trained the proposed architecture using a simulated data set of $10000$ randomly generated rectangular rooms, considering one frequency out of forty in the range \SI[parse-numbers=false]{30-300}{\hertz} for each room.
In particularly, we sample each room into a grid of $32 \times 32$ uniformly-spaced points (independently of the room
dimensions).
Thus, the RTFs are in the form $\mathbf{P} \in \mathbb{R}^{32 \times 32}$.
For each room, we then randomly generated a binary mask $\mathbf{M} \in \{0,1\}^{32 \times 32}$, in order to select the number of microphones placed in the room.
In particular, during training we considered masks selecting a number of microphones in the range $64-512$, which corresponds to the \SI[parse-numbers=false]{6.25-50}{\percent} of the total sound field.
It is worth noting that, as in \cite{saharia2022palette}, we do not directly provide the mask $\mathbf{M}$ as input to the network.
Instead, we fill the masked measures with random Gaussian noise.
Similarly to the training dataset, the test dataset is composed of 250 rooms randomly simulated rooms, considering forty different frequencies in the range \SI[parse-numbers=false]{30-300}{\hertz} for each room. 
RTFs are approximated by using Green’s function, which represents the solution as an infinite summation of room modes in the room \cite{lluis2020sound}.
The room dimensions are randomly sampled, considering a floor area in the range \SI[parse-numbers=false]{20-60}{\meter\squared}, and constant reverberation time $\operatorname{T60} = \SI{0.6}{\second}$.

We trained the model for 1000 epochs, which we empirically verified to sufficient for the employed architecture to converge, using the Adam optimizer with a fixed $5\times 10^{-5}$ learning rate.
During training, a linear noise schedule ranging from $1\mathrm{e}^{-6}$ to $0.01$ is applied over 2000 time-steps.
Similarly, 2000 refinement steps are employed at inference, considering a linear noise schedule from $1\mathrm{e}^{-4}$ to $0.09$.
\subsection{Results}

We assessed the performance of the proposed sound field reconstruction method in terms of normalized mean squared error (NMSE) between ground truth and reconstructed RTFs.
In particular, we computed the NMSE as
\begin{equation}\label{eq:nmse}
    \text{NMSE} = 10\log_{10}\left(
    \frac{1}{N}
    \sum_{i=0}^N \frac{\|
    \hat{P}_i(\mathbf{r},\omega) - P_i(\mathbf{r},\omega)
    \|^2_2}
    {\|P_i(\mathbf{r},\omega)
    \|^2_2}
    \right)
\end{equation}
where $N$ is the number of samples in the testing dataset.
The performances of the method are computed with respect to the number of microphones placed in each room.
Thus, for each room we consider 64, 128, 256, 512 microphones, randomly arranged in space.

Figure~\ref{fig:nmse}(a) shows the NMSE value with respect to frequency, for reconstructions performed using the proposed SF-Diff method and following the setup described in Section~\ref{sec:setup}.
As expected, the number of available measurements highly affects the reconstruction error, which ranges from a minimum value of \SI{-8.35}{\decibel} in the 64 mics configuration, to a minimum value of \SI{-45.39}{\decibel} in the 512 mics configuration.
Having access to more information about the RTF in a certain room, the model is able to better reconstruct the pressure values in the unknown positions.
Also, in all configurations, the error increases with the frequency value.
This is due to the fact that RTFs at higher frequencies present complicated magnitude patterns, graphically, which are more difficult to reconstruct.
In fact, the spatial distribution of room modes becomes more intricate, resulting in peaks and nulls in the RTF at specific locations.
At lower frequencies, instead, the room response is more uniform and thus easier to reconstruct.

Figure~\ref{fig:abs_plots}(c) shows an example of RTF reconstruction performed using SF-Diff, considering the measurement setup represented in Figure~\ref{fig:abs_plots}(a) and ground-truth RTF represented in Figure~\ref{fig:abs_plots}(b).
As it can be seen, the method is able to provide a coherent reconstruction, in which most characteristics of the ground-truth image are present, leading to an NMSE value of \SI{-11.72}{\decibel}.


As a comparison, we performed RTF reconstruction using the methods proposed in \cite{ueno2018kernel} and in \cite{lluis2020sound}, leveraging the same dataset employed in our testing procedure.
Specifically, the approach proposed in \cite{ueno2018kernel} addresses the problem by exploiting the kernel ridge regression with the constraint of the Helmholtz equation. On the other hand, \cite{lluis2020sound} exploits a super-resolution approach based on deep learning.

Results are shown in Figure~\ref{fig:nmse}(b) and Figure~\ref{fig:nmse}(c), respectively. In scenarios where 64 measurements are available, both baseline methods demonstrate superior reconstruction performance, with respect to SF-Diff.
In particular, \cite{ueno2018kernel} achieves a minimum NMSE of \SI{-15.78}{\decibel} at \SI{34}{\hertz}, while \cite{lluis2020sound} obtains a minimum NMSE of \SI{-22.13}{\decibel} at the same frequency.
However, when evaluating setups with an increased number of available measurements (128, 256 and 512 microphones), the reconstruction error does not diminish as prominently as observed in the case of SF-Diff, indicating a limited performance characteristic under varying measurement conditions.

%% file: conclusion.tex
\section{Conclusion}
\label{sec:conclusion}
In this paper we have, to the best of our knowledge, proposed the first application of Denoising Diffusion Probabilistic Models to the problem of sound field reconstruction. Specifically, we consider a pre-existing architecture used for image-inpainting and adapt accordingly. We consider sound fields on a grid positioned on a two-dimensional plane, where measurements are available only for a limited number of arbitrarily positioned microphones. Following the DDPM procedure, we inject noise in the positions for which no measurements available and through the proposed model, we learn to denoise them in order to reconstruct the ground truth sound field. Through an experimental simulation campaign we compare the performance of the proposed technique with a signal processing-based kernel interpolation method and a learning-based method, demonstrating the effectiveness of the DDPM-based approach. The obtained results encourage us to further develop the application of DDPMs to the problem of sound field reconstruction by adapting the technique in order to perform in more challenging scenarios.

%% file: Template.bbl
\begin{thebibliography}{10}

\bibitem{schmid2021spatial}
J.~M. Schmid, E.~Fernandez-Grande, M.~Hahmann, C.~Gurbuz, M.~Eser, and
  S.~Marburg, ``Spatial reconstruction of the sound field in a room in the
  modal frequency range using bayesian inference,'' {\em The Journal of the
  Acoustical Society of America}, vol.~150, no.~6, pp.~4385--4394, 2021.

\bibitem{tylka2020fundamentals}
J.~G. Tylka and E.~Y. Choueiri, ``Fundamentals of a parametric method for
  virtual navigation within an array of ambisonics microphones,'' {\em J. Audio
  Eng. Soc.}, vol.~68, no.~3, pp.~120--137, 2020.

\bibitem{thiergart2013geometry}
O.~Thiergart, G.~Del~Galdo, M.~Taseska, and E.~A.~P. Habets, ``Geometry-based
  spatial sound acquisition using distributed microphone arrays,'' {\em Trans.
  Acoust., Speech, Signal Process.}, vol.~21, no.~12, pp.~2583--2594, 2013.

\bibitem{pezzoli2020parametric}
M.~Pezzoli, F.~Borra, F.~Antonacci, S.~Tubaro, and A.~Sarti, ``A parametric
  approach to virtual miking for sources of arbitrary directivity,'' {\em
  IEEE/ACM Trans. Acoust., Speech, Signal Process.}, vol.~28, pp.~2333--2348,
  2020.

\bibitem{mccormack2022parametric}
L.~McCormack, A.~Politis, R.~Gonzalez, T.~Lokki, and V.~Pulkki, ``Parametric
  ambisonic encoding of arbitrary microphone arrays,'' {\em Trans. Acoust.,
  Speech, Signal Process.}, vol.~30, pp.~2062--2075, 2022.

\bibitem{pezzoli2018reconstruction}
M.~{Pezzoli}, F.~{Borra}, F.~{Antonacci}, A.~{Sarti}, and S.~{Tubaro},
  ``Reconstruction of the virtual microphone signal based on the distributed
  ray space transform,'' in {\em 26th European Signal Processing Conference
  (EUSIPCO)}, pp.~1537--1541, IEEE, 2018.

\bibitem{ueno2017sound}
N.~Ueno, S.~Koyama, and H.~Saruwatari, ``Sound field recording using
  distributed microphones based on harmonic analysis of infinite order,'' {\em
  IEEE Signal Processing Letters}, vol.~25, no.~1, pp.~135--139, 2017.

\bibitem{ueno2018kernel}
N.~Ueno, S.~Koyama, and H.~Saruwatari, ``Kernel ridge regression with
  constraint of helmholtz equation for sound field interpolation,'' in {\em
  Int. Workshop Acoust. Signal Enhanc.}, pp.~1--440, IEEE, 2018.

\bibitem{jin2015theory}
W.~Jin and W.~B. Kleijn, ``Theory and design of multizone soundfield
  reproduction using sparse methods,'' {\em Trans. Acoust., Speech, Signal
  Process.}, vol.~23, no.~12, pp.~2343--2355, 2015.

\bibitem{pezzoli2022sparsity}
M.~Pezzoli, M.~Cobos, F.~Antonacci, and A.~Sarti, ``Sparsity-based sound field
  separation in the spherical harmonics domain,'' in {\em Int. Conf. Acoust.
  Speech Signal Process}, IEEE, 2022.

\bibitem{borra2019soundfield}
F.~Borra, I.~D. Gebru, and D.~Markovic, ``Soundfield reconstruction in
  reverberant environments using higher-order microphones and impulse response
  measurements,'' in {\em Int. Conf. Acoust. Speech Signal Process},
  pp.~281--285, IEEE, 2019.

\bibitem{fahim2017sound}
A.~Fahim, P.~N. Samarasinghe, and T.~D. Abhayapala, ``Sound field separation in
  a mixed acoustic environment using a sparse array of higher order spherical
  microphones,'' in {\em 2017 Hands-free Speech Communications and Microphone
  Arrays (HSCMA)}, pp.~151--155, IEEE, 2017.

\bibitem{koyama2019sparse}
S.~Koyama and L.~Daudet, ``Sparse representation of a spatial sound field in a
  reverberant environment,'' {\em IEEE Journal of Selected Topics in Signal
  Processing}, vol.~13, no.~1, pp.~172--184, 2019.

\bibitem{daminano2021sound}
S.~Damiano, F.~Borra, A.~Bernardini, F.~Antonacci, and A.~Sarti, ``Soundfield
  reconstruction in reverberant rooms based on compressive sensing and
  image-source models of early reflections,'' in {\em 2021 IEEE Workshop on
  Applications of Signal Processing to Audio and Acoustics (WASPAA)}, IEEE,
  2021.

\bibitem{ribeiro2022region}
J.~G. Ribeiro, N.~Ueno, S.~Koyama, and H.~Saruwatari, ``Region-to-region kernel
  interpolation of acoustic transfer functions constrained by physical
  properties,'' {\em Trans. Acoust., Speech, Signal Process.}, vol.~30,
  pp.~2944--2954, 2022.

\bibitem{ribeiro2023kernel}
J.~G. Ribeiro, S.~Koyama, and H.~Saruwatari, ``Kernel interpolation of acoustic
  transfer functions with adaptive kernel for directed and residual
  reverberations,'' {\em arXiv preprint arXiv:2303.03869}, 2023.

\bibitem{bainco2019acousticdeepreview}
M.~J. Bianco, P.~Gerstoft, J.~Traer, E.~Ozanich, M.~A. Roch, S.~Gannot, and
  C.-A. Deledalle, ``Machine learning in acoustics: Theory and applications,''
  {\em The Journal of the Acoustical Society of America (JASA)}, vol.~146,
  no.~5, pp.~3590--3628, 2019.

\bibitem{olivieri2021physics}
M.~Olivieri, M.~Pezzoli, F.~Antonacci, and A.~Sarti, ``A physics-informed
  neural network approach for nearfield acoustic holography,'' {\em Sensors},
  vol.~21, no.~23, 2021.

\bibitem{lluis2020sound}
F.~Llu{\'\i}s, P.~Mart{\'\i}nez-Nuevo, M.~Bo~M{\o}ller, and S.~Ewan~Shepstone,
  ``Sound field reconstruction in rooms: Inpainting meets super-resolution,''
  {\em The Journal of the Acoustical Society of America}, vol.~148, no.~2,
  pp.~649--659, 2020.

\bibitem{pezzoli2022deep}
M.~Pezzoli, D.~Perini, A.~Bernardini, F.~Borra, F.~Antonacci, and A.~Sarti,
  ``Deep prior approach for room impulse response reconstruction,'' {\em
  Sensors}, vol.~22, no.~7, p.~2710, 2022.

\bibitem{ulyanov2018deep}
D.~Ulyanov, A.~Vedaldi, and V.~Lempitsky, ``Deep image prior,'' in {\em
  Proceedings of the IEEE conference on computer vision and pattern
  recognition}, pp.~9446--9454, 2018.

\bibitem{pezzoli2023implicit}
M.~Pezzoli, F.~Antonacci, and A.~Sarti, ``Implicit neural representation with
  physics-informed neural networks for the reconstruction of the early part of
  room impulse responses,'' in {\em Forum Acusticum 2023}, EAA, 2023.

\bibitem{karakonstantis2023generative}
X.~Karakonstantis and E.~Fernandez-Grande, ``Generative adversarial networks
  with physical sound field priors,'' {\em The Journal of the Acoustical
  Society of America}, vol.~154, no.~2, pp.~1226--1238, 2023.

\bibitem{sitzmann2020implicit}
V.~Sitzmann, J.~Martel, A.~Bergman, D.~Lindell, and G.~Wetzstein, ``Implicit
  neural representations with periodic activation functions,'' {\em Advances in
  Neural Information Processing Systems}, vol.~33, pp.~7462--7473, 2020.

\bibitem{fernandez2023generative}
E.~Fernandez-Grande, X.~Karakonstantis, D.~Caviedes-Nozal, and P.~Gerstoft,
  ``Generative models for sound field reconstruction,'' {\em J. Acou. Soc.
  Am.}, vol.~153, no.~2, pp.~1179--1190, 2023.

\bibitem{ho2020denoising}
J.~Ho, A.~Jain, and P.~Abbeel, ``Denoising diffusion probabilistic models,''
  {\em Advances in neural information processing systems}, vol.~33,
  pp.~6840--6851, 2020.

\bibitem{rombach2022high}
R.~Rombach, A.~Blattmann, D.~Lorenz, P.~Esser, and B.~Ommer, ``High-resolution
  image synthesis with latent diffusion models,'' in {\em Proceedings of the
  IEEE/CVF conference on computer vision and pattern recognition},
  pp.~10684--10695, 2022.

\bibitem{moliner2023solving}
E.~Moliner, J.~Lehtinen, and V.~V{\"a}lim{\"a}ki, ``Solving audio inverse
  problems with a diffusion model,'' in {\em Int. Conf. Acoust. Speech Signal
  Process}, pp.~1--5, IEEE, 2023.

\bibitem{yen2023cold}
H.~Yen, F.~G. Germain, G.~Wichern, and J.~Le~Roux, ``Cold diffusion for speech
  enhancement,'' in {\em Int. Conf. Acoust. Speech Signal Process}, pp.~1--5,
  IEEE, 2023.

\bibitem{yu2023conditioning}
C.-Y. Yu, S.-L. Yeh, G.~Fazekas, and H.~Tang, ``Conditioning and sampling in
  variational diffusion models for speech super-resolution,'' in {\em Int.
  Conf. Acoust. Speech Signal Process}, pp.~1--5, IEEE, 2023.

\bibitem{takahashi2023hierarchical}
N.~Takahashi, M.~Kumar, Y.~Mitsufuji, {\em et~al.}, ``Hierarchical diffusion
  models for singing voice neural vocoder,'' in {\em Int. Conf. Acoust. Speech
  Signal Process}, pp.~1--5, IEEE, 2023.

\bibitem{saharia2022palette}
C.~Saharia, W.~Chan, H.~Chang, C.~Lee, J.~Ho, T.~Salimans, D.~Fleet, and
  M.~Norouzi, ``Palette: Image-to-image diffusion models,'' in {\em ACM
  SIGGRAPH 2022 Conference Proceedings}, pp.~1--10, 2022.

\bibitem{zea2019compressed}
E.~Zea, ``Compressed sensing of impulse responses in rooms of unknown
  properties and contents,'' {\em Journal of Sound and Vibration}, vol.~459,
  p.~114871, 2019.

\bibitem{ronneberger2015u}
O.~Ronneberger, P.~Fischer, and T.~Brox, ``U-net: Convolutional networks for
  biomedical image segmentation,'' in {\em Medical Image Computing and
  Computer-Assisted Intervention--MICCAI 2015: 18th Int. Conf., Munich,
  Germany, October 5-9, 2015, Proceedings, Part III 18}, pp.~234--241,
  Springer, 2015.

\bibitem{nichol2021improved}
A.~Q. Nichol and P.~Dhariwal, ``Improved denoising diffusion probabilistic
  models,'' in {\em International Conference on Machine Learning},
  pp.~8162--8171, PMLR, 2021.

\bibitem{saharia2022image}
C.~Saharia, J.~Ho, W.~Chan, T.~Salimans, D.~J. Fleet, and M.~Norouzi, ``Image
  super-resolution via iterative refinement,'' {\em IEEE Transactions on
  Pattern Analysis and Machine Intelligence}, vol.~45, no.~4, pp.~4713--4726,
  2022.

\end{thebibliography}
